\documentclass[aip,reprint]{revtex4-1}
\usepackage{graphicx}
\usepackage{textcomp}
\usepackage{siunitx}
\usepackage{xspace}
\usepackage{mathtools}
\usepackage{color}

\newcommand{\ie}{i.\,e.\xspace}

\newcommand{\eg}{e.\,g.\xspace}

\newcommand{\alo}{Al$_2$O$_3$\xspace}
\newcommand{\sio}{SiO$_2$\xspace}

\begin{document}

%\preprint{}

\title{Highly tuneable hole quantum dots in Ge-Si core-shell nanowires} %Title of paper



\author{Matthias Brauns}
\email[Corresponding author, e-mail: ]{m.brauns@utwente.nl}
\author{Joost Ridderbos}
\affiliation{NanoElectronics Group, MESA+ Institute for Nanotechnology, University of Twente, P.O. Box 217, 7500 AE Enschede, The
Netherlands}
\author{Ang Li}
\affiliation{Department of Applied Physics, Eindhoven University of Technology, Postbox 513, 5600 MB Eindhoven, The Netherlands}
\author{Wilfred G. van der Wiel}
\affiliation{NanoElectronics Group, MESA+ Institute for Nanotechnology, University of Twente, P.O. Box 217, 7500 AE Enschede, The
Netherlands}
\author{Erik P. A. M. Bakkers}
\affiliation{Department of Applied Physics, Eindhoven University of Technology, Postbox 513, 5600 MB Eindhoven, The Netherlands}
\affiliation{QuTech and Kavli Institute of Nanoscience, Delft University of Technology, 2600 GA Delft, The Netherlands}
\author{Floris A. Zwanenburg}
\affiliation{NanoElectronics Group, MESA+ Institute for Nanotechnology, University of Twente, P.O. Box 217, 7500 AE Enschede, The
Netherlands}


\date{\today} 

\begin{abstract}
We define single quantum dots of lengths varying from 60 nanometers up to nearly half a micron in Ge-Si core-shell nanowires. The charging energies scale inversely with the quantum dot length between 18 and 4 meV. Subsequently we split up a long dot into a double quantum dot with separate control over the tunnel couplings and the electrochemical potential of each dot. Both single and double quantum dot configurations prove to be very stable and show excellent control over the electrostatic environment of the dots, making this system a highly versatile platform for spin-based quantum computing.
\end{abstract}


\maketitle %\maketitle must follow title, authors, abstract and \pacs

\section{Introduction}

For spin-based quantum computing \citep{Loss1998}, increasing research efforts have focused in recent years on C, Si, and Ge \citep{Laird2014,Zwanenburg2013,Amato2014} because they can be isotopically enriched to only contain nuclei with zero spin \citep{Itoh2003, Itoh1993} and thus exhibit exceptionally long spin lifetimes \citep{Muhonen2014a,Veldhorst2014}. The one-dimensional character of Ge-Si core-shell nanowires leads to unique electronic properties in the valence band, where heavy and light hole states are mixed \citep{Csontos2007,Csontos2009,Kloeffel2011a}. Early experiments in Ge-Si core-shell nanowires include experiments on double quantum dots\citep{Hu2007} and spin relaxation times.\citep{Hu2012} The band mixing causes an enhanced Rashba-type spin-orbit interaction (SOI) \citep{Kloeffel2011a}, which can be exploited for efficient spin manipulation\citep{Kloeffel2013a}. Therefore Ge-Si core-shell nanowires are an ideal platform for future quantum computation applications. 

In this Letter, we define single quantum dots of several lengths in a Ge-Si core-shell nanowire. We controllably split longer quantum dots up into double quantum dots with tuneable interdot tunnel coupling. Both single and double quantum dots show an exceptional degree of measurement stability.
%\label{}
\section{Device design}\label{sec:sdqd_devices}

\begin{figure}
	\includegraphics[bb=0 0 239 126]{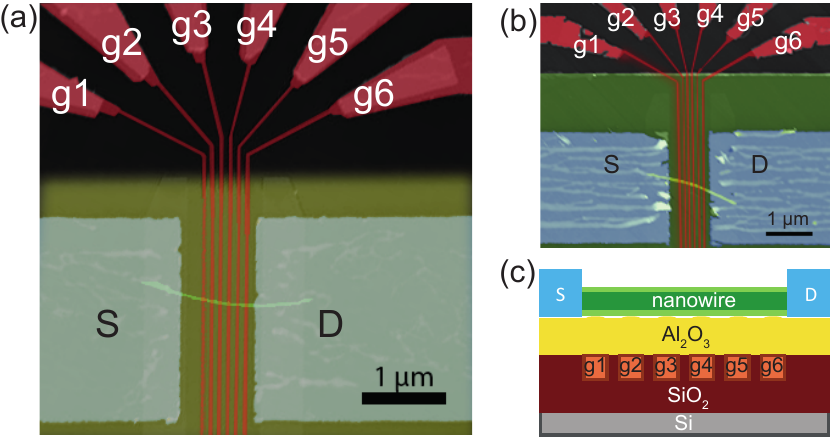}
	\caption{False-colour AFM image of device (a) \emph{D1}, and (b) \emph{D2}. (c) Schematic cross-section displaying the p++-doped Si substrate (grey) with \SI{200}{\nano\meter} of \sio (dark red), six bottom gates g1-g6 (light red), each ~\SI{35}{\nano\meter} wide and with a pitch of \SI{100}{\nano\meter}. The bottom gates are buried under \SI{10}{\nano\meter} of \alo (yellow), on top of which the nanowire is deposited (green) and ohmic contacts (\SI{0.5/50}{\nano\meter} Ti/Pd, blue) are defined.\label{fig:sdqd_device}}	
\end{figure}

We will discuss measurements in two different devices \emph{D1} and \emph{D2} (see Fig.~\ref{fig:sdqd_device}) on two different chips, which have been fabricated in the same way: A p$^\text{++}$-doped Si substrate is covered with \SI{200}{\nano\meter} \sio, on which six bottom gates \emph{g1}-\emph{g6} with \SI{100}{\nano\meter} pitch are patterned with electron beam lithography (EBL). Before metallization of the bottom gates, a 13~s buffered hydrofluoric acid dip etches 20~nm deep trenches into the \sio, so that the bottom gates (approximately 20~nm thick) are sunken into the \sio for an improved planarity. The gates are covered with \SI{10}{\nano\meter} \alo grown with atomic layer deposition at \SI{100}{\celsius}. Two single nanowires  with a Si shell thickness of ~\SI{2.5}{\nano\meter} and a Ge core radius of ~\SI{8}{\nano\meter} (\emph{D1}) and \SI{9}{\nano\meter} (\emph{D2}) are deterministically placed on top of the gate structure with a micromanipulator. Based on transmission electron microscopy studies of similar wires, both the core and the shell are monocrystalline, and their axis is likely pointed along the $<$110$>$ crystal axis.\citep{Li2016} Subsequently we define ohmic contacts to the nanowires and gate contacts made of Ti/Pd (\SI{0.5/50}{\nano\meter}) with EBL. The nanowire parts above the bottom gates are at no point exposed to the electron beam, preventing carbon deposition and introduction of defects into the otherwise defect-free Ge core. All measurements are performed using dc electronic equipment in a dilution refrigerator with a base temperature of 8~mK. A bias voltage $V_\text{SD}$ is applied to source, the current $I$ is measured at the drain contact. An effective hole temperature of $T_\text{hole} \approx \SI{30}{\milli\kelvin}$ has been determined in one of the devices by measuring the temperature dependence of the Coulomb peak width.\citep{Mueller2013,Goldhaber-Gordon1998}

\section{Single quantum dots of varying length}\label{sec:sdqd_sqd}

By using different gates to induce tunnel barriers we can form quantum dots in our nanowire with lengths varying from very long quantum dots (using \emph{g1} and \emph{g6}) to very short dots (using adjacent gates). This flexible quantum dot length together with a tuneable tunnel coupling between the quantum dot and the reservoirs is a great improvement compared to using lateral heterostructures \citep{Fuhrer2007,Roddaro2011a}, or Schottky barriers at the nanowire-metal interface with the contacts \citep{Zwanenburg2009a,Nilsson2011}.

We assume the length of our gate-defined quantum dots to be the the distance between the inner edges of the barrier gates. Using a gate width of $\sim \SI{40}{\nano\meter}$ this results in quantum dot lengths of $\sim \SI{60}{\nano\meter}$ for adjacent barrier gates, $\sim \SI{160}{\nano\meter}$ for barrier gates with one plunger gate in between, $\sim \SI{260}{\nano\meter}$ for two plunger gates, $\sim \SI{360}{\nano\meter}$ for three plunger gates, and $\sim \SI{460}{\nano\meter}$ for four plunger gates, \ie we are able to tune the dot length over almost an order of magnitude.

\begin{figure}
	\includegraphics[bb=0 0 241 363]{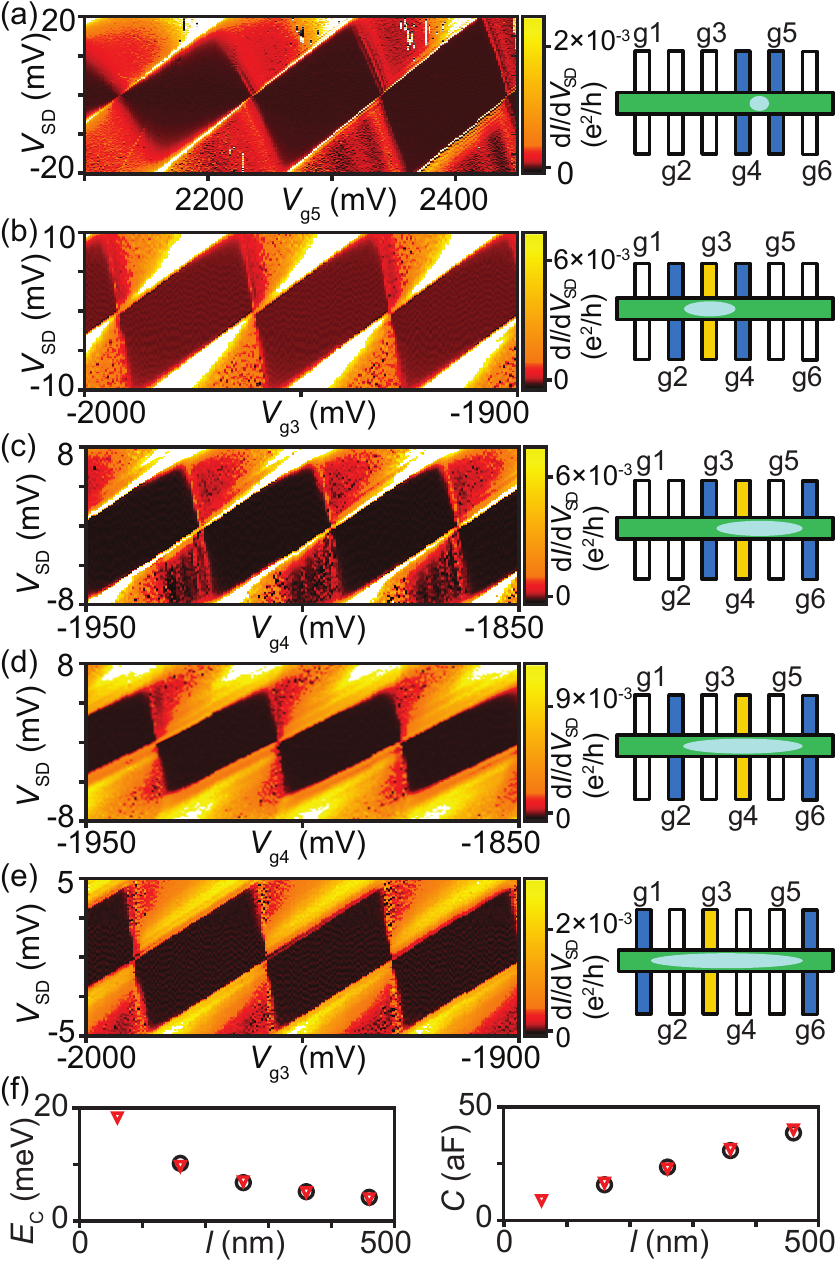}
\caption{Bias spectroscopy of gate-defined single quantum dots formed with (a) zero, (b) one, (c) two, (d) three, and (e) four gates between the barrier gates (indicated in blue). (a) is measured in \emph{D1}, (b)-(e) in \emph{D2}. (f) Charging energy $E_\text{C}$ (left) and total capacitance $C$ (right) of the dot plotted versus the dot length $l$. Red triangles measured in \emph{D1}, black circles in \emph{D2}.}\label{fig:sdqd_sqdlength}
\end{figure}

In Fig.~\ref{fig:sdqd_sqdlength}(a)-(e) we plot $dI/dV \equiv dI/dV_\text{SD}$ versus $V_\text{SD}$ and the voltage on the plunger gate $V_\text{P}$. The formation of quantum dots of five different lengths is reflected in the clear Coulomb diamonds. The shortest quantum dot is formed in device \emph{D1} [Fig.~\ref{fig:sdqd_sqdlength}(a)]. The quantum dots formed with one up to four plunger gates are formed in both devices \emph{D1} and \emph{D2}, Figs.~\ref{fig:sdqd_sqdlength}(b)-(e) display bias spectroscopies of quantum dots formed in \emph{D2}.

%\subsection*{Charging energies}
We extract the respective charging energies $E_\text{C}$ for both devices from the Coulomb diamond height and find a decreasing $E_\text{C}$ from \SI{18.3}{\milli\electronvolt} to \SI{4.2}{\milli\electronvolt}, inversely proportional to the increasing dot length [see Fig.~\ref{fig:sdqd_sqdlength}(f) and Tab.~\ref{tab:sqd_parameters}]. Since $E_\text{C}$ is linked to the total capacitance $C$ of the quantum dot via  $E_\text{C} = e^2/C$,\cite{Kouwenhoven1997} $C$ is directly proportional to the quantum dot length. $E_\text{C}$ and $C$ are highly consistent for the two devices.

For the quantum dot configurations with a dedicated plunger gate [Figs.~\ref{fig:sdqd_sqdlength}(b)-(e)], $E_\text{C}$ as well as the shape of the Coulomb diamonds stay constant over several charge transitions, reflecting the validity of the constant interaction model. In Fig.~\ref{fig:sdqd_sqdlength}(a), $E_\text{C}$ increases significantly from \SI{16.7}{\milli\electronvolt} to \SI{20.0}{\milli\electronvolt} and also the slopes of the Coulomb diamond edges change. We attribute this to using the right barrier gate as a plunger, leading to a decreasing dot size and changing capacitive couplings to this barrier gate and the adjacent reservoir. Therefore, the constant interaction model is not valid in this configuration, and the tuneability of the quantum dot is limited compared to the longer quantum dots with a dedicated plunger gate. We extract the values for $E_\text{C}$ and $C$ for the zero-plunger configuration from the middle Coulomb diamond, for which they are in line with those for the longer dots. 

\begin{table}
	\caption{\label{tab:sqd_parameters}Parameters for electrostatically defined quantum dots of varying length as extracted from Fig.~\ref{fig:sdqd_sqdlength}(a)-(e).}
	\begin{ruledtabular}
	\begin{tabular}{rrrrrr}
		%\tableheadline{line} & 
		%$I_\text{pos}(\epsilon=0)$\newline (\si{\pico\ampere}) & 
		{$l$ (\si{\nano\meter})} & 
		{$E_\text{C}$ (\si{\milli\electronvolt})} &
		{$C$ (\si{\atto\farad})} &
		{$\Delta V_\text{g}$ (\si{\milli\volt})} &
		{$C_\text{g}$ (\si{\atto\farad})} \\ \hline
    	60 	& \num{18.3(2)}	& \num{8.8(2)}	& \num{104(1)} 	& \num{1.54(2)} \\
    	160 & \num{10.2(2)}	& \num{15.7(3)}	& \num{31.5(2)}	& \num{5.09(3)} \\
    	260 & \num{6.8(2)}	& \num{23.5(5)}	& \num{29.6(4)}	& \num{5.41(7)} \\
    	360 & \num{5.2(1)} 	& \num{30.8(6)}	& \num{28.6(4)}	& \num{5.63(8)} \\
    	460 & \num{4.2(1)}	& \num{38.6(9)}	& \num{29.7(4)}	& \num{5.39(7)} \\
	\end{tabular}
	\end{ruledtabular}
\end{table}

%\subsection*{Capacitances}
The constant charging energies over several Coulomb diamonds in Figs.~\ref{fig:sdqd_sqdlength}(b)-(e) are accompanied by constant Coulomb peak spacings $\Delta V_\text{g}$ at $V_\text{SD} = 0$, indicating a constant gate capacitance $C_\text{g}$ over several charge transitions, another indication for the validity of the constant interaction model. If we now compare the plunger gate capacitances between Fig.~\ref{fig:sdqd_sqdlength}(b) and (e), we find them to be all very similar, ($\sim \SI{5.5}{\atto\farad}$), while the total capacitance increases linearly by $\sim \SI{7.5}{\atto\farad}$ per additional plunger gate [see Fig~\ref{fig:sdqd_sqdlength}(f)]. The discrepancy of \SI{\sim2}{\atto\farad} can be explained by the finite capacitance of the global back gate which increases with the dot length and the change in the self-capacitance of the quantum dot. The linearly increasing total capacitance indicates equal coupling of all gates, consistent with the gate geometry (equal width and distance to the nanowire). In Fig~\ref{fig:sdqd_sqdlength}(f) we also plot $E_\text{C}$ and $C$ for quantum dots formed in \emph{D1} with at least one plunger gate alongside the data for \emph{D2}. The consistency between the data and therefore demonstrates a high degree of control over the electrostatic environment of the gate-defined quantum dot. 
%Note that $E_C$ for our longest quantum dots is approximately the same as in \citet{}

% \begin{figure}
% 	\includegraphics{figs/fig_3.pdf}
% \caption{a) Single bias spectroscopy measurement (top five panels) of gate-defined single quantum dot formed with barrier gates \emph{g2} and \emph{g4} (similar to Fig.~\ref{fig:sdqd_sqdlength}b) in \emph{D2}. Distance of adjacent Coulomb peaks for 63 consecutive Coulomb diamonds (lower panel).}\label{fig:sdqd_1plunger_hugerange}
% \end{figure}
\section{Tuneable double quantum dots}\label{sec:sdqd_dqd}

Tuneable double quantum dots are essential for spin readout via Pauli spin blockade\citep{Ono2002}. For a fully tuneable double quantum dot we need five gates: Three barrier gates to form tunnel barriers, and two plunger gates to tune the electrochemical potential of each dot separately. We use device \emph{D1} starting from a situation equivalent to Fig.~\ref{fig:sdqd_sqdlength}(d), and increase the voltage on the middle gate $V_{g4}$. When approaching the pinch-off voltage, a tunnel barrier is formed and the single quantum dot splits up into two tunnel-coupled quantum dots.

\begin{figure}
	\includegraphics[bb=0 0 241 377]{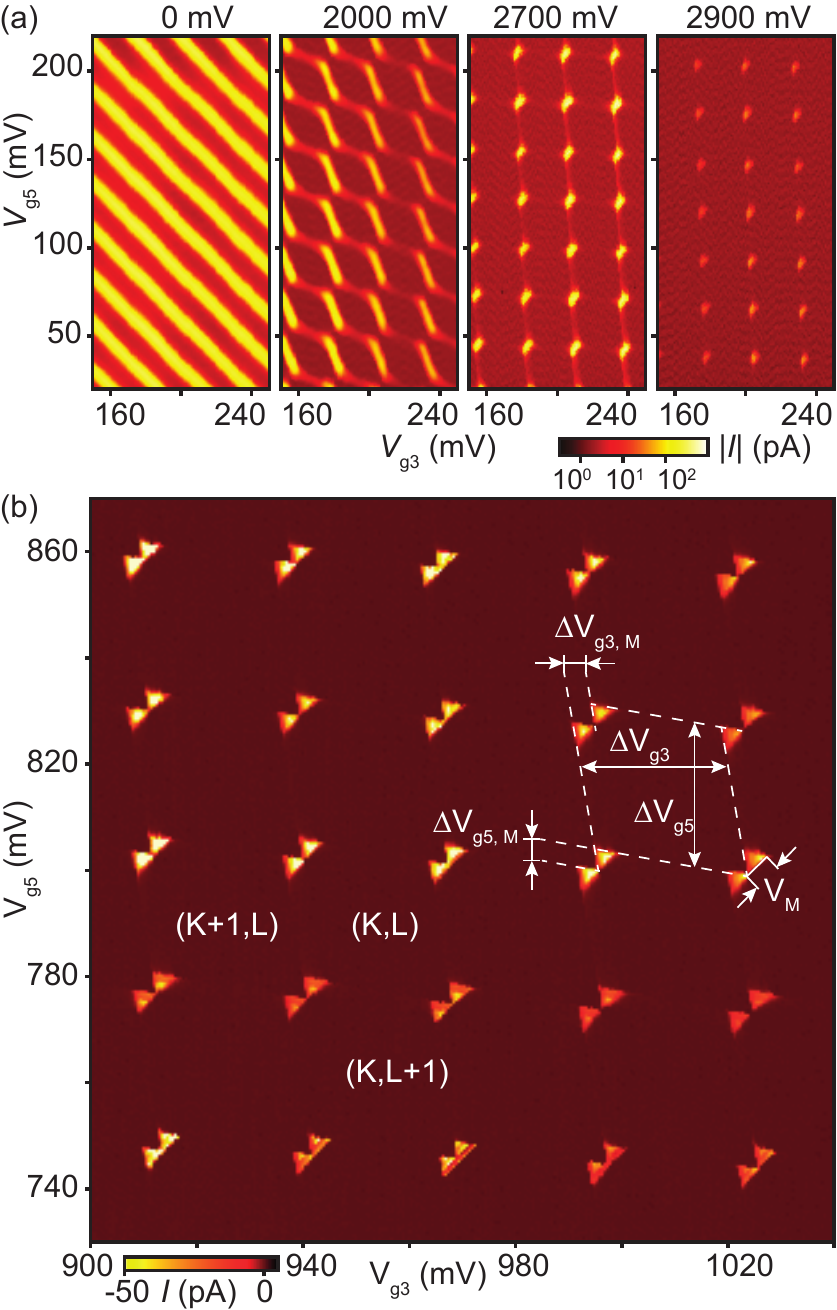}
	\caption{(a) Charge stability diagrams with current $I$ plotted versus $V_{g3}$ and $V_{g5}$ for varying voltages on \emph{g4} at fixed $V_{g2} = \SI{2490}{\milli\volt}$ and $V_{g6} = \SI{2940}{\milli\volt}$. (b) Charge stability diagram of a double quantum dot at $V_{g2} = \SI{2500}{\milli\volt}$, $V_{g4} = \SI{2100}{\milli\volt}$, and $V_{g6} = \SI{3180}{\milli\volt}$. $(k,l)$ denote the charge occupation numbers on the left ($k$) and right dot ($l$). All measurements performed on device \emph{D1}.\label{fig:sdqd_dqd}}
\end{figure}

The charge stability diagrams at four different $V_{g4}$ are plotted in Fig.~\ref{fig:sdqd_dqd}(a). We keep the outer barrier gates at constant voltages ($V_{g2} = \SI{2490}{\milli\volt}$, $V_{g6} = \SI{2940}{\milli\volt}$), and plot the current at a fixed $V_\text{SD} = \SI{1}{\milli\volt}$. For $V_{g4} = \SI{0}{\milli\volt}$ we observe the typical stability diagram of a single quantum dot.\citep{Wiel2002} The spacing of the diagonal, parallel lines of finite current along the respective plunger gate axis is directly related to the capacitance between the quantum dot and this gate: $C_g = e/\Delta V_g$. We observe $\Delta V_g \approx \SI{27}{\milli\volt}$ for both \emph{g3} and \emph{g5}, i.e. both gates have the same capacitance $C_g \approx \SI{5.8}{\atto\farad}$ to the quantum dot. This indicates that the quantum dot indeed stretches over the whole distance between the tunnel barriers above gates \emph{g2} and \emph{g6} and is also in agreement with the gate capacitances in Table~\ref{tab:sqd_parameters}. 

At $V_{g4} = \SI{2000}{\milli\volt}$ the straight lines have evolved into a regular honeycomb pattern with two distinct slopes that form the long edges of each honeycomb, indicating the formation of a strongly coupled double quantum dot.\cite{Wiel2002} From the distance between adjacent parallel lines we extract the voltages needed to add a hole to the left (right) dot $\Delta V_{g3}$, ($\Delta V_{g5}$) and calculate the corresponding capacitances. For the left dot we find $\Delta V_{g3} = \SI{25.1(5)}{\milli\volt}$ and for the right dot $\Delta V_{g5} = \SI{26.1(5)}{\milli\volt}$, resulting in gate capacitances $C_{g3} = \SI{6.3(2)}{\atto\farad}$ and $C_{g5} = \SI{6.1(2)}{\atto\farad}$. The sets of honeycomb edges representing the addition of a hole to either the left or the right dot are both significantly slanted because of the mutual capacitive coupling $C_M$ between the two dots that leads to a separation between the two triple points.\citep{Wiel2002} We express this shift in terms of gate voltages and find $\Delta V_{g3,M} = \SI{9.2(5)}{\milli\volt}$ and $\Delta V_{g5,M} = \SI{10.9(5)}{\milli\volt}$. Using the expression $C_{g,M} = C_{g} \Delta V_{g,M} / \Delta V_{g}$,\citep{Wiel2002} we obtain $C_{g3,M} = \SI{2.3(3)}{\atto\farad}$ and $C_{g5,M} = \SI{2.5(3)}{\atto\farad}$. A second mechanism affecting the slopes of the honeycomb edges is the finite cross capacitance between \emph{g3} and the right dot $C_{g3,C}$, and \emph{g5} and the left dot $C_{g5,C}$. This cross capacitance leads to a shift of the triple points along the \emph{g3} gate axis while changing the charge occupation of the right dot, and along the \emph{g5} gate axis while changing the number of holes on the left dot. This effect is very weak, and we extract $C_{g3,C} \approx C_{g5,C} \approx \SI{0.1}{\atto\farad}$.

Increasing the voltage on the interdot barrier gate to $V_{g4} = \SI{2700}{\milli\volt}$ only slightly changes the gate capacitances to $C_{g3} = \SI{5.9(2)}{\atto\farad}$ and $C_{g5} = \SI{5.7(2)}{\atto\farad}$. For the mutual gate capacitances we find a much stronger relative change to $C_{g3,M} = \SI{0.6(1)}{\atto\farad}$ and $C_{g5,M} = \SI{0.7(1)}{\atto\farad}$, which indicates a significantly increased separation of the charge distribution of both dots. The now only faintly visible long edges of the honeycombs also suggest a decreased tunnel coupling to the reservoirs so that cotunnelling is suppressed \citep{DeFranceschi2001}. A finite, but very small cross capacitance of the plunger gates is also observed here, again on the order of \SI{0.1}{\atto\farad}.

A further increase of the interdot barrier gate to $V_{g4} = \SI{2900}{\milli\volt}$ completely quenches the cotunnelling current at the long honeycomb edges, so that now transport is only possible at the triple point pairs. This indicates well defined charge states confined in the quantum dots weakly coupled to the reservoirs. Again we observe a slight decrease of the gate capacitances to $C_{g3} = \SI{5.8(2)}{\atto\farad}$ and $C_{g5} = \SI{5.5(2)}{\atto\farad}$, and also the mutual capacitances decrease further to $C_{g3,M} = \SI{0.4(1)}{\atto\farad}$ and $C_{g5} = \SI{0.4(1)}{\atto\farad}$. All extracted capacitances are summarized in Table~\ref{tab:sdqd_dqdformation}.

\begin{table}
	\caption{\label{tab:sdqd_dqdformation}Capacitances for increasing voltage on the middle barrier gate \emph{g4} of an electrostatically defined single ($V_{g4} = \SI{0}{\milli\volt}$) or double quantum dot ($V_{g4} \ge \SI{2000}{\milli\volt}$) as extracted from Fig.~\ref{fig:sdqd_dqd}(a).}
	\begin{ruledtabular}
	\begin{tabular}{rrrrr} 
		{$V_{g4}$ (\si{\milli\volt})} & 
		{$C_{g3}$ (\si{\atto\farad})} &
		{$C_{g5}$ (\si{\atto\farad})} &
		{$C_{g3,M}$ (\si{\atto\farad})} &
		{$C_{g5,M}$ (\si{\atto\farad})} \\ \hline
    	0 		& \num{5.8(3)}	& \num{5.8(3)}	&				&  \\
    	2000	& \num{6.3(2)}	& \num{6.1(2)}	& \num{2.3(3)}	& \num{2.5(3)} \\
    	2700 	& \num{5.9(2)}	& \num{5.7(2)}	& \num{0.6(2)}	& \num{0.7(2)} \\
    	2900 	& \num{5.8(2)} 	& \num{5.5(2)}	& \num{0.4(1)}	& \num{0.4(1)} \\
	\end{tabular}
	\end{ruledtabular}
\end{table}

In Fig.~\ref{fig:sdqd_dqd}(b) we show a high-re\-so\-lu\-tion stability diagram of a double quantum dot wealky coupled to the reservoirs at $V_\text{SD} = \SI{-1.5}{\milli\volt}$ with barrier gate voltages of $V_{g2} = \SI{2500}{\milli\volt}$, $V_{g4} = \SI{2100}{\milli\volt}$, and $V_{g6} = \SI{3180}{\milli\volt}$. Clearly visible is a very regular pattern of 25 bias triangle pairs, from which we extract the gate-to-dot capacitances in the same way as before. We obtain $C_{g3} = \SI{5.9(2)}{\atto\farad}$ and $C_{g5} = \SI{5.9(2)}{\atto\farad}$, and mutual capacitances of $C_{\text{M,}g3} = \SI{0.9(1)}{\atto\farad}$ and $C_{\text{M,}g5} = \SI{0.9(1)}{\atto\farad}$. The increased values for $C_{\text{M},i}$ indicate an indeed increased capacitive coupling between the dots. 

We extract the charging energies from Fig.~\ref{fig:sdqd_dqd}(b) by relating the bias triangle size to an energy of \SI{1.5}{\milli\electronvolt}. We obtain a charging energy of the left dot $U_1 = \SI{10.6(5)}{\milli\electronvolt}$ and of the right dot  $U_2 = \SI{9.3(5)}{\milli\electronvolt}$. For the mutual charging energy $U_\text{M}$ we extract $U_\text{M} = \SI{1.5(2)}{\milli\electronvolt}$. The size and shape of the bias triangles are exceptionally stable over the whole range of the measurement. This underlines the high degree of control over the electrochemical potentials of the quantum dots as well as the tunnel and capacitive couplings. Our devices are therefore exceptionally suitable for direct-transport experiments in comparison to other systems, where tunnel couplings change strongly when changing the charge occupation.\citep{Fasth2005a,Liu2008a,Sand-Jespersen2008,Li2015a} Such experiments are relevant, because the applied bias between the two reservoirs serves as an energy scale, which, \eg, allows for the determination of the singlet-triplet splitting\citep{Brauns2016} and the Zeeman splitting.\citep{Brauns2015}

In summary we demonstrate a high degree of control over the charge distribution in a double quantum dot. We have changed the mutual capacitances, a measure for the degree of separation of the dots, by a factor of six while keeping the capacitances  between the left (right) dot and \emph{g3} (\emph{g5}) almost constant. The corresponding charging energies are in agreement with the experiments on single quantum dots of the same length in \emph{D2}.

\section{Conclusion}\label{sec:sdqd_conclusion}

In conclusion, we have electrostatically formed highly-tunable single and double quantum dots inside Ge-Si core-shell nanowires. We can vary the length of the single quantum dots from \SI{60}{\nano\meter} to \SI{460}{\nano\meter} corresponding to charging energies of the quantum dots varying from $\sim \SI{18}{\milli\electronvolt}$ down to $\sim \SI{4}{\milli\electronvolt}$. 

Furthermore, we have split a single quantum dot into a double quantum dot in a controlled way. Our low-cross-capacitance gate design enables us to keep the voltage on the outer barriers constant while varying the interdot barrier, \ie it is not necessary to retune all gates. All capacitances and charging energies extracted from single and double quantum dot measurements are highly consistent. 25 bias triangle pairs form a very regular pattern in the stability diagram with constant triangle sizes, indicating an exceptional degree of control over the tunnel couplings over a large range of gate voltages. 

This combination of tuneability and stability makes Ge-Si core-shell nanowires an ideal platform for further experiments towards quantum computation applications.

\begin{acknowledgments}
We thank Sergey Amitonov, and Paul-Christiaan Spruijtenburg for fruitful discussions. We acknowledge technical support by Hans Mertens. FAZ acknowledges financial support through the EC FP7-ICT initiative under Project SiAM No 610637, and from the Foundation for Fundamental Research on Matter (FOM), which is part of the Netherlands Organization for Scientific Research (NWO). EPAMB acknowledges financial support through the EC FP7-ICT initiative under Project SiSpin No 323841.
\end{acknowledgments}


\end{document}